\documentstyle[aps,pra,twocolumn,psfig]{revtex}

\begin{document}

\title{Logical dissipation of automata implements -
Dissipation of computation}
\author{Philippe Matherat$^a$ and Marc-Thierry Jaekel$^b$ \\
$^a$ Ecole Nationale Sup\'erieure des T\'el\'ecommunications, CNRS,\\
46, rue Barrault, F75013 Paris\\
$^b$ Laboratoire de Physique Th\'eorique de l'Ecole Normale Su\-p\'e\-rieu\-re,
CNRS, UPS\\
24 rue Lhomond F75231 Paris Cedex 05}
\date{
LPTENS 98/17A)\\
English translation of:\\
{\it Dissipation logique des impl\'ementations d'automates -
Dissipation du calcul}\\
(Technique et Science Informatiques {\bf 15} (1996) 1079
}
\maketitle
\begin{abstract}
\end{abstract}
As revealed by discussions of principle on energy dissipation by computers,
logic imposes constraints on physical systems
designed for a logical function.
We define a notion of logical dissipation for a finite automaton.
We discuss the constraints associated with physical implementation of automata
and exhibit the role played by modularity for testability. As a result,
practical computers, which are necessarily modular, dissipate
proportionally to computation time.


\section{Introduction}

Finite automata are mathematical objects, while the notion of dissipation
has its origin in the second principle of thermodynamics and refers to physical
systems.
On one hand, automata are defined in a logical context where
states are {\it successive} in the sense of natural numbers
and are not successive in time.
On another hand, the evolution of an arbitrary physical system cannot be
reduced
to transitions between the states of a finite automaton.
Nevertheless, implementation
of automata under the form of physical systems meets constraints which
logic imposes on physics, and conversely.

A physical system, whether it exists spontaneously or it has been set up by an
experimentalist, is a set of elements which obey dynamical laws.
Its evolution can be described by initial conditions and the laws of motion.
The possibility for the evolution to take the form of a finite set of
states related by transitions is not left to the physicist's choice.
However, in a great number of practical situations, the physical system
under study is not imposed on the experimentalist but, on the contrary,
is designed, built and verified by the experimentalist himself.
When one designs a machine, one chooses the part which will be played
by macroscopic degrees of freedom. If one chooses to let a piston
move within two extreme positions, one will set the necessary material
to insure proper steering, and one will {\it a priori} reject all
products which would not insure a satisfactory rigidity or would suffer
from too early weariness. Similarly, a designer of electronic
logical devices looking for the implementation of an automaton
will work to impose the logical function and will reject all arrangements
which do not show sufficient reliability.

The question of the {\it existence} of a physical implement realising
a given logical function, in a reliable and stable manner, has no evident
answer. It leads to the question of the proof of a physical device.
One needs to get convinced that the physical object will effectively
realise the assigned function in all configurations of use and will
continue to realise it in the future. This is in general guaranteed by a
{\it test} which verifies that the physical object behaves as its
logical definition requests.

Let us admit that such implements of automata do exist and that they
do not reduce to ordinary physical systems. We shall try to exhibit
their characteristic properties.

To the category of machines realising a function chosen by the experimentalist
belong many physicist's instruments, many measurement devices.
The fundamental laws of physics are observed by means of devices
whose functions which have been assigned must be describable in
logical terms. Thus, even if physical systems cannot be reduced to
automata, experimentation on physical systems necessarily involves
automata.

We shall not raise here the general question of the relation between
the structure of logical language and the form of physical laws.
We shall rather study the constraints which are imposed (by their logical
function) on physical systems which are implements of a finite and
deterministic automaton or of a deterministic Turing machine,
particularly in relation with the question of dissipation.

\subsection{History}

We very briefly recall here the essential steps which underlined the relation
between computation and dissipation. A detailed bibliography can be found in
\cite{LEF 90}.

The first example of an automaton taking part in a physical process was
introduced by Maxwell, in relation with the statistical character
of the second principle of thermodynamics \cite{MAX 75}.

Taking over the analysis of Maxwell demon, Szilard made it precise
that such a physical device, which memorizes information, must also
dissipate, otherwise it would allow to build a perpetual motion of
the second kind \cite{SZI 29}. By relying on the second principle
of thermodynamics, he computed that a one-bit memory, hence a two-state
automaton, such that its two memory states correspond to entropy
increases $S_1$ and $S_2$ must satisfy:
$$e^{-{S_1\over k}} + e^{-{S_2\over k}} \leq 1$$
with $k$ Boltzmann constant ($1.38$ $10^{-23}$ Joules/Kelvin).
This result is established by reasoning over a complete cycle of a
mono-thermal machine made of a perfect gas and one memory.

Some authors attributed to {\it measurement} (initialisation of the memory)
the origin of this dissipation (Gabor \cite{GAB 51}, Brillouin \cite{BRI 59}),
while Landauer preferred to locate the latter in the {\it erasure}
(forgetting of previously memorized information) \cite{LAN 61}.
Although both interpretations seem opposite we shall show in section V
that this is not the case for finite automata.

Reinterpreting Szilard, Brillouin suggests a generalised Carnot principle
reading $\Delta (S - I) \geq 0$ where $S$ stands for entropy and $I$
for information, both in the same unit. In the opposite way, Landauer
suggests that a system which memorizes one bit of information at temperature
$T$ must dissipate an energy of the order of at least $kT \rm{ln} 2$
for each bit erasure, whatever its physical realisation (mechanical,
electronic, etc...) \cite{LAN 61}. This point of view insists on the
general character of the {\it minimal} dissipation, related to the
logical function, even if the effectively dissipated energy may depend
on choices which have been made in the physical realisation of the
memory. This "Landauer principle" (as was called later by Bennett)
focuses on {\it logical irreversibilities}:
a device which does not insure logical reversibility should not pretend
to physical reversibility.

These two principles (Brillouin and Landauer) thus shift, each in its way,
the discussion of dissipation,
from the domain of physics to the domain of logics.
They suggest to use a notion of {\it logical dissipation}, measured by the
quantity of lost information, which could then be transposed to the domain
of physics under the form of a minimal energy dissipation.

Landauer priniciple, which we shall rather call {\it criterion},
has never been confronted to experiments, due to the smallness
of the invoked dissipated energy. It is, at ordinary temperature,
smaller by many orders of magnitude than the energy effectively dissipated
by the best devices presently existing, which leads to think that
"Landauer dissipation" is screened by more important dissipative effects
whose origin is still not well understood. Discussions of principle
also show that its expression can only be approximate, and must in
particular be questioned at very low temperatures (see for
instance \cite{LIK 82}).

Following Landauer, Bennett then showed that with any Turing machine $\cal S$
one can associate a Turing machine $\cal R$, which performs
the same computations as $\cal S$, but which furthermore memorizes
the history of the
computation, then erases this history, the whole being done in a logically
reversible way. He deduced that any computation can be performed with a
dissipation which is not related to the number of steps in the computation,
but to the length of the result only \cite{BEN 73}.

Here, we shall not discuss the validity or the physical evaluation
of Landauer criterion, but we shall study the properties of
logical dissipation. Although defined logically, this dissipation
characterises physical systems. Thus, we shall show that the constraints
on physical realisations of information systems induced by the necessity
of testability impose particular logical structures which result in their
logical dissipation. We shall see that Bennett's construction, even if it
permits the logical reversibility of any computation, does not provide
a non-dissipative implement that one can exhibit explicitly
before the beginning of the computation (which does not depend on
the particular computation).
For universal reversible machines, the constraints associated with
physical implementation lead to dissipative realisations.

\subsection{Plan}

The notions of {\it finite automaton} and {\it logical dissipation}
should first be made precise. The existence of this last notion is implicitly
assumed in Brillouin and Landauer considerations, but no definition has
been given up to now.

We shall discuss the physical implementation of automata. Contrarily
to the automaton which is defined formally, its physical realisation
must satisfy further constraints so that to ensure its logical function.
It should in particular allow to check that the physical object
effectively performs its function. We shall restrict ourselves to physical
objects whose function can be tested and does not evolve in time.

We shall then introduce the notion of {\it modularity} for an
automaton implementation.
Modular structures allow to considerably simplify the
test of implements and thus to realise automata of
a very high complexity. We shall show that a counterpart is that these
implementations lead to dissipation.

Finally, we shall discuss the possibilities of physical implementations
of deterministic Turing machines and their dissipation. We shall show that
the implements of computing machines dissipate proportionally to
computation time.

\section{Finite automata}

Finite automata can be used in various contexts \cite{PER 95}.
When they are used to define a language, their states are not comparable
with the states of a physical system. On another hand, when they are used to
describe a sequential electronic device, a state of the
mathematical automaton corresponds to a stable physical state of the
electronic device, and transitions of the mathematical automaton
correspond to physical transitions which are ordered in physical time.
In such a situation, every transition is a validation of the preceding
input chain which has led to this state. It is this second point
of view that we shall adopt.

Our definition of an automaton will remain very close to classical definitions,
but it shall differ on certain points, in order to allow for a definition of
the notions of {\it convergence} and {\it divergence}. We shall try to
use the ordinary vocabulary as much as possible
(see for instance \cite{MIN 67} or
\cite{STE 90}).

We shall see the automaton as a {\it black box} (figure 1).
Inputs are stimuli ($S$) from the exterior world. Outputs are responses
($R$) directed to the exterior world. The different allowed values for $S$
are in finite number (symbols chosen in a finite alphabet).
And similarly for $R$ (the alphabets are in general distinct, i.e.
with different numbers of symbols for $S$ and $R$).

\bigskip
\begin{center}
\begin{picture}(0,0)%
\psfig{file=ldiaf1.pstex}%
\end{picture}%
\setlength{\unitlength}{0.00087500in}%
\begingroup\makeatletter\ifx\SetFigFont\undefined
\def\x#1#2#3#4#5#6#7\relax{\def\x{#1#2#3#4#5#6}}%
\expandafter\x\fmtname xxxxxx\relax \def\y{splain}%
\ifx\x\y   
\gdef\SetFigFont#1#2#3{%
  \ifnum #1<17\tiny\else \ifnum #1<20\small\else
  \ifnum #1<24\normalsize\else \ifnum #1<29\large\else
  \ifnum #1<34\Large\else \ifnum #1<41\LARGE\else
     \huge\fi\fi\fi\fi\fi\fi
  \csname #3\endcsname}%
\else
\gdef\SetFigFont#1#2#3{\begingroup
  \count@#1\relax \ifnum 25<\count@\count@25\fi
  \def\x{\endgroup\@setsize\SetFigFont{#2pt}}%
  \expandafter\x
    \csname \romannumeral\the\count@ pt\expandafter\endcsname
    \csname @\romannumeral\the\count@ pt\endcsname
  \csname #3\endcsname}%
\fi
\fi\endgroup
\begin{picture}(2667,1194)(631,-973)
\put(1441,-781){\makebox(0,0)[lb]{\smash{\SetFigFont{12}{14.4}{rm}State $Q$}}}
\put(1441,-106){\makebox(0,0)[lb]{\smash{\SetFigFont{12}{14.4}{rm}Automaton:}}}
\put(1441,-331){\makebox(0,0)[lb]{\smash{\SetFigFont{12}{14.4}{rm}Black box}}}
\put(2836, 74){\makebox(0,0)[lb]{\smash{\SetFigFont{12}{14.4}{rm}Output}}}
\put(631, 74){\makebox(0,0)[lb]{\smash{\SetFigFont{12}{14.4}{rm}Input}}}
\put(811,-421){\makebox(0,0)[lb]{\smash{\SetFigFont{12}{14.4}{rm}$S$}}}
\put(2836,-421){\makebox(0,0)[lb]{\smash{\SetFigFont{12}{14.4}{rm}$R$}}}
\end{picture}

\bigskip
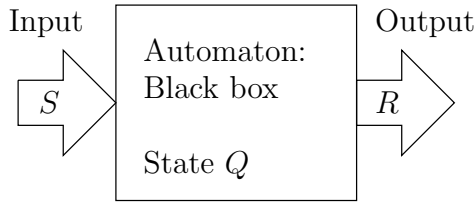
{\bf Figure 1.} {\it Finite automaton}
\end{center}

\bigskip
To denote the successive values of the internal state $Q$ of the automaton,
and the successive values of inputs and outputs alike, we shall use an
index $t$ (time) taking integer values. For the moment,
this is not the physical time, as succession refers to natural numbers only.
When physical implements will be defined, the index $t$ will become
a discrete physical time (rhythmed by a clock exterior to the automaton).

The automaton evolution is described by the two following
functions $F$ and $G$:
$$Q_{t+1} = G(Q_t, S_t)$$
$$R_t = F(Q_t)$$
where $Q_t$, $S_t$ and $R_t$ denote the state, the input stimuli and the output
responses of the automaton, at time $t$. $G$ is called the transition function.

We shall use a graphic representation for $G$, which associates a circle
with each state and an arrow with each state transition (figure 2), and
which we shall call the {\it state diagram} or the {\it graph of the
automaton}.

\bigskip
\begin{center}
\begin{picture}(0,0)%
\psfig{file=ldiaf2.pstex}%
\end{picture}%
\setlength{\unitlength}{0.00087500in}%
\begingroup\makeatletter\ifx\SetFigFont\undefined
\def\x#1#2#3#4#5#6#7\relax{\def\x{#1#2#3#4#5#6}}%
\expandafter\x\fmtname xxxxxx\relax \def\y{splain}%
\ifx\x\y   
\gdef\SetFigFont#1#2#3{%
  \ifnum #1<17\tiny\else \ifnum #1<20\small\else
  \ifnum #1<24\normalsize\else \ifnum #1<29\large\else
  \ifnum #1<34\Large\else \ifnum #1<41\LARGE\else
     \huge\fi\fi\fi\fi\fi\fi
  \csname #3\endcsname}%
\else
\gdef\SetFigFont#1#2#3{\begingroup
  \count@#1\relax \ifnum 25<\count@\count@25\fi
  \def\x{\endgroup\@setsize\SetFigFont{#2pt}}%
  \expandafter\x
    \csname \romannumeral\the\count@ pt\expandafter\endcsname
    \csname @\romannumeral\the\count@ pt\endcsname
  \csname #3\endcsname}%
\fi
\fi\endgroup
\begin{picture}(1221,430)(804,-743)
\put(1171,-421){\makebox(0,0)[lb]{\smash{\SetFigFont{10}{12.0}{rm}$S$}}}
\put(901,-646){\makebox(0,0)[lb]{\smash{\SetFigFont{10}{12.0}{rm}$Q$}}}
\put(1801,-646){\makebox(0,0)[lb]{\smash{\SetFigFont{10}{12.0}{rm}$Q'$}}}
\end{picture}

\bigskip
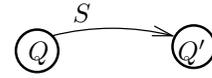
{\bf Figure 2.} {\it A transition}
\end{center}

\bigskip

By requiring that $G$ be a function we shall restrict ourselves to
{\it deterministic} automata.

In our definition of function $F$, the response $R_t$ only depends
on the state $Q_t$. Consequently, a change of input can only
modify the output at the following time, and {\it through} a change of state.
This definition means that all information available on the output
lies in the state of the automaton. This we shall call a {\it sequential}
function.

For simplicity, $F$ will be required to be injective. The exterior world
will know the instantaneous state of the automaton through the instantaneous
value of the response. This will simplify the test of a physical implement
of the automaton.

The chain of input symbols will be called {\it input word}, and the ouput
 chain {\it ouput word} alike. Their length can be infinite if the state
diagram
contains a loop (allows to return to a state already visited).

\subsection{Divergence and convergence}

First of all, the arrows leaving a same state will be imposed to
arrive at different states. If two different symbols $A$ and $B$
trigger a same transition from state $Q$ to state $Q'$, then
this transition will be said to be triggered by label "$A or B$",
and only one arrow will appear on the state diagram (figure 3).
Thus, state diagrams will not be multigraphs, but oriented and labelled graphs.

\bigskip
\begin{center}
\begin{picture}(0,0)%
\psfig{file=ldiaf3.pstex}%
\end{picture}%
\setlength{\unitlength}{0.00087500in}%
\begingroup\makeatletter\ifx\SetFigFont\undefined
\def\x#1#2#3#4#5#6#7\relax{\def\x{#1#2#3#4#5#6}}%
\expandafter\x\fmtname xxxxxx\relax \def\y{splain}%
\ifx\x\y   
\gdef\SetFigFont#1#2#3{%
  \ifnum #1<17\tiny\else \ifnum #1<20\small\else
  \ifnum #1<24\normalsize\else \ifnum #1<29\large\else
  \ifnum #1<34\Large\else \ifnum #1<41\LARGE\else
     \huge\fi\fi\fi\fi\fi\fi
  \csname #3\endcsname}%
\else
\gdef\SetFigFont#1#2#3{\begingroup
  \count@#1\relax \ifnum 25<\count@\count@25\fi
  \def\x{\endgroup\@setsize\SetFigFont{#2pt}}%
  \expandafter\x
    \csname \romannumeral\the\count@ pt\expandafter\endcsname
    \csname @\romannumeral\the\count@ pt\endcsname
  \csname #3\endcsname}%
\fi
\fi\endgroup
\begin{picture}(3256,1664)(578,-1523)
\put(2701,-286){\makebox(0,0)[lb]{\smash{\SetFigFont{10}{12.0}{rm}$A or B$}}}
\put(3466,-376){\makebox(0,0)[lb]{\smash{\SetFigFont{10}{12.0}{rm}$Q'$}}}
\put(2836,-646){\makebox(0,0)[lb]{\smash{\SetFigFont{10}{12.0}{rm}$Q$}}}
\put(1486,-376){\makebox(0,0)[lb]{\smash{\SetFigFont{10}{12.0}{rm}$Q'$}}}
\put(856,-646){\makebox(0,0)[lb]{\smash{\SetFigFont{10}{12.0}{rm}$Q$}}}
\put(1171,-826){\makebox(0,0)[lb]{\smash{\SetFigFont{10}{12.0}{rm}$B$}}}
\put(946,-331){\makebox(0,0)[lb]{\smash{\SetFigFont{10}{12.0}{rm}$A$}}}
\end{picture}

\bigskip
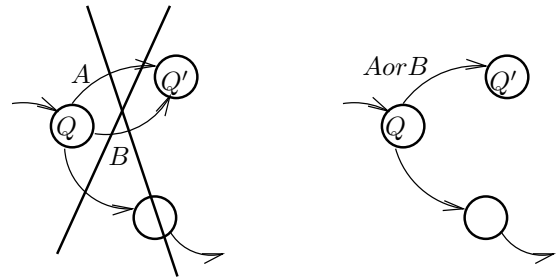
{\bf Figure 3.} {\it Only one arrow from state $Q$ to state $Q'$}
\end{center}

\bigskip
{\bf Definition}: when at least two transitions start from a
same state (thus corresponding
to different input values and ending at different states), we shall say
that there is a {\bf divergence}.

Remark: in absence of divergence, there are two possibilities:

- {\it indifferent} input: the environment provides some input,
but the automaton makes the same transition whatever the value of this input;
the transition will be labelled by the set of all possible input values.

- {\it implicit} input: the automaton waits for a particular input value,
and the environment can only provide this value.

In following discussions, both cases will be equivalent.

{\bf Definition}: when at least two transitions end at the same state, we shall
say
that there is a {\bf convergence}.

{\bf Definition}: an automaton is said to be {\bf reversible}
if and only if its graph
does not contain any convergence.

Remark: equivalence between reversibility and absence of convergence is only
compatible with a function $F$ defined such that a change of response
only occurs as a consequence of a change of state. All information
on the {\it past evolution} of the automaton lies in its internal state.
If the state is reached through a convergence, some information on the
past evolution of the automaton is lost.

To compare with definitions used in \cite{STE 90}, we shall consider that:

- all states are {\it acceptant}: this property allows logical states to become
stable physical states. All performed transitions are {\it successful}.

- there is no {\it reject state}:
the transition function is only a partial function,
which means that for some states some symbols may not trigger a transition.
In this case, such symbol will be forbidden as input at that time.
Definitions of automata which involve a reject state
lead to a number of arrows
leaving each state which may be higher.
Thus, more states may be sources of divergences,
a situation which we shall not allow.

The automaton is compatible with a class of environments only,
those which provide as stimuli only words which are compatible with a
path in the graph of the automaton ({\it a computation}).
The definition of the automaton entails the definition
of the class of compatible environments. For physical implements
of automata, this will lead to exclude physical environments which are
not compatible with the definition of the automaton.

\section{Physical implementation of automata}

We wish to establish a correspondence between the states of the logical
automaton and stable states of a physical system, and to make the transitions
of the logical automaton correspond to physical transitions.

We shall exclude automata which are defined by a recursive formula
and whose finite character may thus be undecidable (the eventuality of such
a situation will be discussed in section VIII). Moreover, only automata
will be considered whose definition leads to a graph which is stable
in time.

The "black box" of figure 1 now becomes a physical object. Input and output
become channels for communicating with the environment. Time becomes physical.
Physical time being not discrete, we shall assume here that it can be
discretised by means of an external clock, which sends pulses on
an additional input channel of the automaton ($CK$ on figure 4).
Stimuli must also be assumed to be synchronised with the clock, a further
constraint imposed by the automaton on its environment. The clock pulses
trigger internal transformations of the automaton and we shall assume
that after some delay, shorter than a clock period, the physical state
becomes stable, in such a manner that it can represent one of the logical
states of the automaton. This does not require a physical state which
does not evolve, but entails a definite correspondence with a single logical
state.

\bigskip
\begin{center}
\begin{picture}(0,0)%
\psfig{file=ldiaf4.pstex}%
\end{picture}%
\setlength{\unitlength}{0.00087500in}%
\begingroup\makeatletter\ifx\SetFigFont\undefined
\def\x#1#2#3#4#5#6#7\relax{\def\x{#1#2#3#4#5#6}}%
\expandafter\x\fmtname xxxxxx\relax \def\y{splain}%
\ifx\x\y   
\gdef\SetFigFont#1#2#3{%
  \ifnum #1<17\tiny\else \ifnum #1<20\small\else
  \ifnum #1<24\normalsize\else \ifnum #1<29\large\else
  \ifnum #1<34\Large\else \ifnum #1<41\LARGE\else
     \huge\fi\fi\fi\fi\fi\fi
  \csname #3\endcsname}%
\else
\gdef\SetFigFont#1#2#3{\begingroup
  \count@#1\relax \ifnum 25<\count@\count@25\fi
  \def\x{\endgroup\@setsize\SetFigFont{#2pt}}%
  \expandafter\x
    \csname \romannumeral\the\count@ pt\expandafter\endcsname
    \csname @\romannumeral\the\count@ pt\endcsname
  \csname #3\endcsname}%
\fi
\fi\endgroup
\begin{picture}(2757,1824)(541,-1603)
\put(541,-1231){\makebox(0,0)[lb]{\smash{\SetFigFont{10}{12.0}{rm}synchronised}}
}
\put(541,-1411){\makebox(0,0)[lb]{\smash{\SetFigFont{10}{12.0}{rm}on $CK$}}}
\put(541,-1051){\makebox(0,0)[lb]{\smash{\SetFigFont{10}{12.0}{rm}Input}}}
\put(811,-421){\makebox(0,0)[lb]{\smash{\SetFigFont{10}{12.0}{rm}$S$}}}
\put(1441,-331){\makebox(0,0)[lb]{\smash{\SetFigFont{10}{12.0}{rm}Black box}}}
\put(1441,-781){\makebox(0,0)[lb]{\smash{\SetFigFont{10}{12.0}{rm}State $Q$}}}
\put(2836,-421){\makebox(0,0)[lb]{\smash{\SetFigFont{10}{12.0}{rm}$R$}}}
\put(2341,-1546){\makebox(0,0)[lb]{\smash{\SetFigFont{10}{12.0}{rm}$CK$}}}
\put(2836,-916){\makebox(0,0)[lb]{\smash{\SetFigFont{10}{12.0}{rm}Output}}}
\put(1441,-151){\makebox(0,0)[lb]{\smash{\SetFigFont{10}{12.0}{rm}Automaton:}}}
\end{picture}

\bigskip
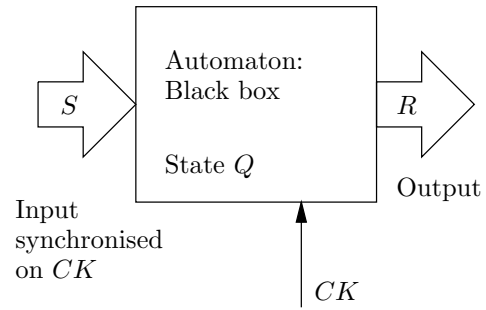
{\bf Figure 4.} {\it Physical implement of an automaton}
\end{center}

\bigskip

\subsection{Existence}

Such a physical system is not an object to be investigated once its existence
has been recognised, but on the contrary an object which has been described
by its function before any knowledge of its existence, of its compatibility
with fundamental properties of physical systems, which may impose further
limitations.

We shall assume that such physical implements of automata do exist
(taking possible constraints of implementation into account) and shall study
the consequences of our definitions. If only a reduced class of physical
objects does comply with the definitions and implementation constraints,
we shall know that these consequences apply to this reduced class only.

We shall be guided by the existence of electronic logical devices
(mainly computing machines) which do behave as such objects, with the
usual constraints of implementation and use: on electric power supply,
temperature, allowed values for inputs,
fan-in and fan-out, clock synchronisation of
inputs, etc... These constraints are quoted only to indicate what may determine
the existence, without providing any proof of existence, and none of these
constraints will be considered as necessary.

\subsection{Test of a physical implement}

In order to give physical existence to a logical function,
the designer looks for an implementation which corresponds to the
specification conditions described by the graph,
and then guarantees the logical
functioning of the implement he has found, for any use which remains within
limits fixed by contract. Conformity of the implement to the graph
is determined by a {\it test}.

By test, one must understand that the physical object is activated by
a machine presenting all the input symbols which are necessary to explore
the whole graph, that is to visit all states and to follow all transitions.
Only automata for which this is possible will be considered.

Such notion of test is related to the stability of the function in time.
It forbids any evolution of the graph during functioning. The state of the
automaton changes, but its graph cannot. One assumes that a future functioning
in other environments can be deduced from a correct functioning at a given
time on a test machine. This leads to introduce a notion of {\it date of test}.

The test is actually necessary to define the physical implement of
the automaton. It is indeed out of question to ensure the functioning
of the automaton by a dynamic description of all its constituents.
The test allows to associate a {\it characteristic graph} to a physical object,
without worrying any more about the internal physical behavior of the object.

\subsection{Cost of an implement}

The {\it cost} of an implement will be defined by the complexity, in
computation time, of the test program. The test must check the
transition function, and thus all transitions. The number of transitions is
smaller or equal to the product of the number of states by the number
of input symbols. For a given vocabulary, the number of transitions
thus grows like the number of states. To test a particular transition,
it may be necessary to follow a part of the graph. At worst, the cost thus
grows like the square of the number of states.

\section{Maxwell demons}

Maxwell demons are examples of physical systems which behave as automata.
One-bit memories, like those used by Szilard and Landauer, are physical
implements of two-states automata like the one of figure 5.

\bigskip
\begin{center}
\begin{picture}(0,0)%
\psfig{file=ldiaf5.pstex}%
\end{picture}%
\setlength{\unitlength}{0.00087500in}%
\begingroup\makeatletter\ifx\SetFigFont\undefined
\def\x#1#2#3#4#5#6#7\relax{\def\x{#1#2#3#4#5#6}}%
\expandafter\x\fmtname xxxxxx\relax \def\y{splain}%
\ifx\x\y   
\gdef\SetFigFont#1#2#3{%
  \ifnum #1<17\tiny\else \ifnum #1<20\small\else
  \ifnum #1<24\normalsize\else \ifnum #1<29\large\else
  \ifnum #1<34\Large\else \ifnum #1<41\LARGE\else
     \huge\fi\fi\fi\fi\fi\fi
  \csname #3\endcsname}%
\else
\gdef\SetFigFont#1#2#3{\begingroup
  \count@#1\relax \ifnum 25<\count@\count@25\fi
  \def\x{\endgroup\@setsize\SetFigFont{#2pt}}%
  \expandafter\x
    \csname \romannumeral\the\count@ pt\expandafter\endcsname
    \csname @\romannumeral\the\count@ pt\endcsname
  \csname #3\endcsname}%
\fi
\fi\endgroup
\begin{picture}(2511,1375)(766,-1461)
\put(2296,-1276){\makebox(0,0)[lb]{\smash{\SetFigFont{10}{12.0}{rm}set to 0}}}
\put(3016,-736){\makebox(0,0)[lb]{\smash{\SetFigFont{10}{12.0}{rm}set to 1}}}
\put(766,-1096){\makebox(0,0)[lb]{\smash{\SetFigFont{10}{12.0}{rm}set to 0}}}
\put(1756,-376){\makebox(0,0)[lb]{\smash{\SetFigFont{10}{12.0}{rm}set to 1}}}
\put(1763,-1070){\makebox(0,0)[lb]{\smash{\SetFigFont{10}{12.0}{rm}0}}}
\put(2705,-526){\makebox(0,0)[lb]{\smash{\SetFigFont{10}{12.0}{rm}1}}}
\end{picture}

\bigskip
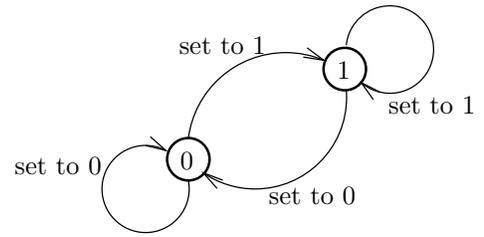
{\bf Figure 5.} {\it One-bit memory}
\end{center}

\bigskip

On each state do converge two arrows, so that the automaton is not reversible.
Every transition is both the registration of one bit ({\it initialisation}
or {\it measurement}) and the {\it erasure} of the bit previously memorised.

Such a memory allows to realise a Maxwell demon, as shown on figure 6, which
has
been derived from Bennett's interpretation of Szilard demon \cite{BEN 87}.
This figure represents a cylinder, closed by two pistons which trap
a molecule, the whole being at a uniform temperature.

\bigskip
\begin{picture}(0,0)%
\psfig{file=ldiaf6.pstex}%
\end{picture}%
\setlength{\unitlength}{0.00087500in}%
\begingroup\makeatletter\ifx\SetFigFont\undefined
\def\x#1#2#3#4#5#6#7\relax{\def\x{#1#2#3#4#5#6}}%
\expandafter\x\fmtname xxxxxx\relax \def\y{splain}%
\ifx\x\y   
\gdef\SetFigFont#1#2#3{%
  \ifnum #1<17\tiny\else \ifnum #1<20\small\else
  \ifnum #1<24\normalsize\else \ifnum #1<29\large\else
  \ifnum #1<34\Large\else \ifnum #1<41\LARGE\else
     \huge\fi\fi\fi\fi\fi\fi
  \csname #3\endcsname}%
\else
\gdef\SetFigFont#1#2#3{\begingroup
  \count@#1\relax \ifnum 25<\count@\count@25\fi
  \def\x{\endgroup\@setsize\SetFigFont{#2pt}}%
  \expandafter\x
    \csname \romannumeral\the\count@ pt\expandafter\endcsname
    \csname @\romannumeral\the\count@ pt\endcsname
  \csname #3\endcsname}%
\fi
\fi\endgroup
\begin{picture}(3021,4434)(619,-4033)
\put(2656,-136){\makebox(0,0)[lb]{\smash{\SetFigFont{6}{7.2}{rm}of the
preceding cycle. Here,}}}
\put(2656,-256){\makebox(0,0)[lb]{\smash{\SetFigFont{6}{7.2}{rm}X means one of
the two states}}}
\put(2656,-376){\makebox(0,0)[lb]{\smash{\SetFigFont{6}{7.2}{rm}indifferently.}}}
\put(2656,-16){\makebox(0,0)[lb]{\smash{\SetFigFont{6}{7.2}{rm}The memory
contains the value}}}
\put(2656,-1006){\makebox(0,0)[lb]{\smash{\SetFigFont{6}{7.2}{rm}The separating
wall is lowered,}}}
\put(2656,-1126){\makebox(0,0)[lb]{\smash{\SetFigFont{6}{7.2}{rm}then the side
where the molecule}}}
\put(2656,-1246){\makebox(0,0)[lb]{\smash{\SetFigFont{6}{7.2}{rm}is located is
registered in the}}}
\put(2656,-1366){\makebox(0,0)[lb]{\smash{\SetFigFont{6}{7.2}{rm}memory (R for
right, L for left).}}}
\put(2656,-1486){\makebox(0,0)[lb]{\smash{\SetFigFont{6}{7.2}{rm}Registration
of the new value in}}}
\put(2656,-1606){\makebox(0,0)[lb]{\smash{\SetFigFont{6}{7.2}{rm}the memory
erases the old one.}}}
\put(2656,-2131){\makebox(0,0)[lb]{\smash{\SetFigFont{6}{7.2}{rm}The piston is
pushed on the side}}}
\put(2656,-2251){\makebox(0,0)[lb]{\smash{\SetFigFont{6}{7.2}{rm}which is
opposite to the one}}}
\put(2656,-2491){\makebox(0,0)[lb]{\smash{\SetFigFont{6}{7.2}{rm}work is
exchanged.}}}
\put(2656,-2371){\makebox(0,0)[lb]{\smash{\SetFigFont{6}{7.2}{rm}indicated by
the memory: no}}}
\put(2656,-3031){\makebox(0,0)[lb]{\smash{\SetFigFont{6}{7.2}{rm}The separating
wall is raised.}}}
\put(2656,-3151){\makebox(0,0)[lb]{\smash{\SetFigFont{6}{7.2}{rm}The molecule
furnishes work to}}}
\put(2656,-3391){\makebox(0,0)[lb]{\smash{\SetFigFont{6}{7.2}{rm}energy from
the (monothermal)}}}
\put(2656,-3751){\makebox(0,0)[lb]{\smash{\SetFigFont{6}{7.2}{rm}Then, one
comes back to the}}}
\put(2656,-3871){\makebox(0,0)[lb]{\smash{\SetFigFont{6}{7.2}{rm}beginning of
the cycle}}}
\put(2656,-3271){\makebox(0,0)[lb]{\smash{\SetFigFont{6}{7.2}{rm}the piston and
restores its kinetic}}}
\put(2656,-3511){\makebox(0,0)[lb]{\smash{\SetFigFont{6}{7.2}{rm}heat
source.}}}
\put(811,299){\makebox(0,0)[lb]{\smash{\SetFigFont{6}{7.2}{rm}Two states}}}
\put(811,179){\makebox(0,0)[lb]{\smash{\SetFigFont{6}{7.2}{rm}memory}}}
\put(1351, 74){\makebox(0,0)[lb]{\smash{\SetFigFont{6}{7.2}{rm}X}}}
\put(1351,-961){\makebox(0,0)[lb]{\smash{\SetFigFont{6}{7.2}{rm}R}}}
\put(1351,-1996){\makebox(0,0)[lb]{\smash{\SetFigFont{6}{7.2}{rm}R}}}
\put(1351,-3031){\makebox(0,0)[lb]{\smash{\SetFigFont{6}{7.2}{rm}R}}}
\put(1621,299){\makebox(0,0)[lb]{\smash{\SetFigFont{6}{7.2}{rm}Movable
separating wall}}}
\put(1756,-3976){\makebox(0,0)[lb]{\smash{\SetFigFont{6}{7.2}{rm}Heat}}}
\end{picture}

\bigskip
\begin{center}
{\bf Figure 6.} {\it Szilard demon}
\end{center}

\bigskip

One must recall that a simple physical system (without memory), like a trap
which would be activated by the passage of the molecule \cite{SMO 12},
could not perform as a Maxwell demon \cite{FEY 63}. To act in this
manner, it is necessary to register the molecule's position in order to
use this information at a later time (in figure 6, the third and fourth steps
depend on the second one). Let us note that measurement, memorization and
erasure are inseparable and closely linked to the treatment of information
specified by the graph of the memory. The logical function of the automaton
characterises and summarizes the additional role played by an observer,
like Maxwell demon, by comparison to a simple physical system.
Szilard's argument shows that, according to the second principle of
thermodynamics, the automaton must dissipate energy in the course of its
functioning.

Possibilities of memorization and loss of information cannot be reduced to
graphs as simple as the one of figure 5. Convergences are not always
associated with symmetrical divergences. Nonetheless, we show in next section
how convergences and divergences allow to characterise the automaton
dissipation.
\section{Logical dissipation}

A notion of logical dissipation, which depends on the graph of the implement
but not on particular physical characteristics, can be defined.

Let us consider as an example an automaton which is a little more complex
than the one-bit memory, as represented by figure 7. Let us also consider the
following two words, of ten binary symbols each,
which are recognised by this automaton:

word 1: 0100001010

word 2: 0011100110

These two words correspond to paths in the graph (two computations) which
can be distinguished by choices made when leaving states B, C and G.

\bigskip
\begin{center}
\begin{picture}(0,0)%
\psfig{file=ldiaf7.pstex}%
\end{picture}%
\setlength{\unitlength}{0.00087500in}%
\begingroup\makeatletter\ifx\SetFigFont\undefined
\def\x#1#2#3#4#5#6#7\relax{\def\x{#1#2#3#4#5#6}}%
\expandafter\x\fmtname xxxxxx\relax \def\y{splain}%
\ifx\x\y   
\gdef\SetFigFont#1#2#3{%
  \ifnum #1<17\tiny\else \ifnum #1<20\small\else
  \ifnum #1<24\normalsize\else \ifnum #1<29\large\else
  \ifnum #1<34\Large\else \ifnum #1<41\LARGE\else
     \huge\fi\fi\fi\fi\fi\fi
  \csname #3\endcsname}%
\else
\gdef\SetFigFont#1#2#3{\begingroup
  \count@#1\relax \ifnum 25<\count@\count@25\fi
  \def\x{\endgroup\@setsize\SetFigFont{#2pt}}%
  \expandafter\x
    \csname \romannumeral\the\count@ pt\expandafter\endcsname
    \csname @\romannumeral\the\count@ pt\endcsname
  \csname #3\endcsname}%
\fi
\fi\endgroup
\begin{picture}(3022,2113)(631,-1958)
\put(721,-1366){\makebox(0,0)[lb]{\smash{\SetFigFont{10}{12.0}{rm}Stop}}}
\put(631,-511){\makebox(0,0)[lb]{\smash{\SetFigFont{10}{12.0}{rm}Start}}}
\put(1351,-1906){\makebox(0,0)[lb]{\smash{\SetFigFont{10}{12.0}{rm}0}}}
\put(1531,-1456){\makebox(0,0)[lb]{\smash{\SetFigFont{10}{12.0}{rm}1}}}
\put(1396,-511){\makebox(0,0)[lb]{\smash{\SetFigFont{10}{12.0}{rm}0}}}
\put(2161,-106){\makebox(0,0)[lb]{\smash{\SetFigFont{10}{12.0}{rm}1}}}
\put(2026,-556){\makebox(0,0)[lb]{\smash{\SetFigFont{10}{12.0}{rm}0}}}
\put(3016,-736){\makebox(0,0)[lb]{\smash{\SetFigFont{10}{12.0}{rm}1}}}
\put(3376,-466){\makebox(0,0)[lb]{\smash{\SetFigFont{10}{12.0}{rm}0}}}
\put(2296,-1366){\makebox(0,0)[lb]{\smash{\SetFigFont{10}{12.0}{rm}1}}}
\put(3331,-1726){\makebox(0,0)[lb]{\smash{\SetFigFont{10}{12.0}{rm}0}}}
\put(2206,-1816){\makebox(0,0)[lb]{\smash{\SetFigFont{10}{12.0}{rm}1}}}
\put(1261,-826){\makebox(0,0)[lb]{\smash{\SetFigFont{10}{12.0}{rm}A}}}
\put(1891,-241){\makebox(0,0)[lb]{\smash{\SetFigFont{10}{12.0}{rm}B}}}
\put(2926,-331){\makebox(0,0)[lb]{\smash{\SetFigFont{10}{12.0}{rm}C}}}
\put(3466,-1321){\makebox(0,0)[lb]{\smash{\SetFigFont{10}{12.0}{rm}D}}}
\put(2341,-1096){\makebox(0,0)[lb]{\smash{\SetFigFont{10}{12.0}{rm}E}}}
\put(2521,-1861){\makebox(0,0)[lb]{\smash{\SetFigFont{10}{12.0}{rm}F}}}
\put(1531,-1771){\makebox(0,0)[lb]{\smash{\SetFigFont{10}{12.0}{rm}G}}}
\end{picture}

\bigskip
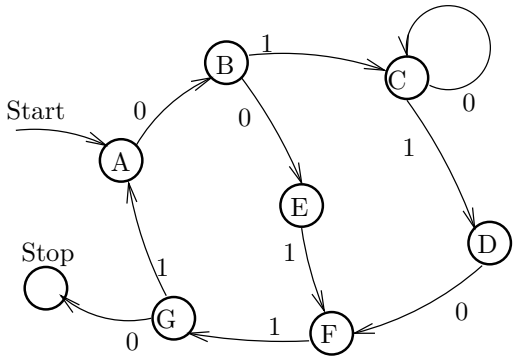
{\bf Figure 7.} {\it Example of a dissipative automaton}
\end{center}

\bigskip

In order to leave one of these three states, the automaton uses the information
provided by the input symbol. One can quantify this information by means of
the probabilities of input symbols, that is $p_{0i}$ and $p_{1i}$ the
probabilities to leave state $i$ by reading $0$ or $1$ (with
$p_{0i} + p_{1i} = 1$). The necessary information for leaving state $i$, as
measured in bits, is given by \cite{SHA 49}:
$$-p_{0i}{\rm log}_2  p_{0i} - p_{1i} {\rm log}_2  p_{1i}$$

This expression is equal to one bit if $p_{0i} = p_{1i}$. It is equal to zero
if the input is indifferent or implicit, that is, if only one
arrow leaves the state in question (in our example, we have considered that
divergenceless transitions are associated with implicit inputs).
Words 1 and 2 correspond to the following paths (or computations):

word 1: Start A {\bf B C C C C C} D F {\bf G} Stop (7 bits of choice)

word 2: Start A {\bf B} E F {\bf G} A {\bf B} E F {\bf G} Stop
(4 bits of choice)

Bold letters (B, C and G) correspond to states which are left using an input
symbol carrying one bit of information (considering both input values as
equiprobable), and the quantity of information carried by the
input word corresponding to the computation is indicated in parentheses.

The divergent paths when leaving states B, C and G are determined by the
information provided by the input word. Thus, when the automaton is in one
of the states C, D or E, it {\it keeps track} of the way it has left state
B. However, once it has reached or left state F, this information is lost.
In the spirit of Landauer criterion, the automaton should dissipate because
of this convergence on state F.

{\bf Definition}:
the {\bf logical dissipation} (or {\it dissipated information})
is the amount of information which is lost at {\it convergences}. The amount
of logical dissipation is evaluated on the {\it divergences} of the graph.

The logical dissipation depends on the particular input word within the
class of words which are recognised by the automaton. Its amount is not
necessarily of one bit by binary symbol. One must consider the probability
of occurence of this particular word among all possible words. If this
set is made of $N$ equiprobable words, the corresponding information is
${\rm log}_2 N$ bits. If the set of words is infinite, the probability
for one word among words of the same length can be considered.
Such a definition of logical dissipation, as the amount of information
carried by a chain of symbols (or word), agrees with Shannon's \cite{SHA 49}
and Brillouin's \cite{BRI 59} results.

The logical dissipation depends on the automaton.
If the automaton is specialised
for a particular output word $r$ (with no divergence), then the amount of
information of the input word $s$ will vanish and all the necessary information
to generate $r$ will be contained in the structure of the graph of the
automaton. Such an automaton does not dissipate. On the contrary, if the
automaton is like the one of figure 5, then the quantity of information carried
by input word $s$ will be equal to the length of the ouput word $r$ (one bit
for each emitted binary symbol). The automaton of figure 5 thus has a structure
which results in memorizing and shifting the word one step further.
Such an automaton dissipates one bit for each binary symbol read in input.

One may consider that the difficulty in relating
the amount of input information
to dissipation (divergences to convergences) is at the origin of differences
of interpretation for the source of dissipation in Szilard's experiment
(the measurement or the erasure). Considering finite automata
allows to conciliate both interpretations.
Measurement and erasure can be opposed
only if one considers local characteristics of the graph,
and not the whole graph. Indeed, the graph of figure 5 could be equivalent
to an infinite binary tree if the automaton were not imposed to be finite.

In the following, {\it dissipation} will be used for logical dissipation
and will be measured in bits.

Automata involving loops will essentially be considered. In such a case, when
part of the computation (of the path) is such that the initial and final states
are identical, the logical dissipation is exactly equal to the information
contained in the corresponding part of the input word. All information
read in input is lost.

\section{Modular implementation of automata}

As seen previously, the cost of the physical implement of an automaton, as
defined by the complexity of the test, grows as the number of transitions
of the automaton, which itself grows as the number of states (or at worst
as its square). Yet, the number of states of automata used in practice can be
enormous. For instance, the number of states of an $n$-bit memory is equal to
$2^n$, while $n$ is commonly larger than $10^6$ in existing computers.
The time required for testing such a function largely exceeds the age of
the Universe ($\sim 3.10^{17}$ seconds), even if the elementary period
of the test program is made as small as Planck time ($\sim 5.10^{-44}$ second).
One can only envisage to implement such functions if their validation
is obtained from tests of the parts they are made of. But this then
requires the existence of a stable function for these parts. We shall call
{\it modular} an implementation which allows to test the parts separately.
Before coming to a more precise definition, we first give an example
which shows in particular that modularity results from a choice.

\subsection{Example of a choice of implementation: modulo four counter}

Let us assume that one wishes to realise the cyclic modulo four counter defined
by the graph of figure 8.

\bigskip
\begin{center}
\begin{picture}(0,0)%
\psfig{file=ldiaf8.pstex}%
\end{picture}%
\setlength{\unitlength}{0.00087500in}%
\begingroup\makeatletter\ifx\SetFigFont\undefined
\def\x#1#2#3#4#5#6#7\relax{\def\x{#1#2#3#4#5#6}}%
\expandafter\x\fmtname xxxxxx\relax \def\y{splain}%
\ifx\x\y   
\gdef\SetFigFont#1#2#3{%
  \ifnum #1<17\tiny\else \ifnum #1<20\small\else
  \ifnum #1<24\normalsize\else \ifnum #1<29\large\else
  \ifnum #1<34\Large\else \ifnum #1<41\LARGE\else
     \huge\fi\fi\fi\fi\fi\fi
  \csname #3\endcsname}%
\else
\gdef\SetFigFont#1#2#3{\begingroup
  \count@#1\relax \ifnum 25<\count@\count@25\fi
  \def\x{\endgroup\@setsize\SetFigFont{#2pt}}%
  \expandafter\x
    \csname \romannumeral\the\count@ pt\expandafter\endcsname
    \csname @\romannumeral\the\count@ pt\endcsname
  \csname #3\endcsname}%
\fi
\fi\endgroup
\begin{picture}(1364,1409)(1254,-1778)
\put(1351,-781){\makebox(0,0)[lb]{\smash{\SetFigFont{10}{12.0}{rm}0}}}
\put(2206,-556){\makebox(0,0)[lb]{\smash{\SetFigFont{10}{12.0}{rm}1}}}
\put(2431,-1546){\makebox(0,0)[lb]{\smash{\SetFigFont{10}{12.0}{rm}2}}}
\put(1441,-1681){\makebox(0,0)[lb]{\smash{\SetFigFont{10}{12.0}{rm}3}}}
\end{picture}

\bigskip
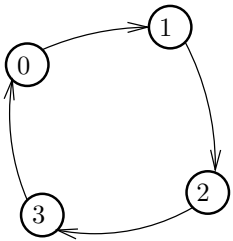
{\bf Figure 8.} {\it Modulo four counter}
\end{center}

\bigskip

Counters of this kind are used to realise frequency dividers.
Such an automaton does not have
any information input, but changes its state at each clock pulse.
Its graph does not contain any divergence nor convergence. Its logical
dissipation vanishes and one can imagine to realise it with a
physical dissipation as small as wanted. Such a physical realisation could
be provided by a rotating wheel, where four sectors representing the four
states would be printed. One easily conceives that such an implementation is
only faced with a dissipation related with friction, which could be diminished
with improved technology and which is not linked to the logical function.

On another hand, a method frequently used by electronic engineers consists
in encoding the state on two binary digits, memorized in one-bit memories.
Let us describe an example of such an implementation.

Let us first define as an automaton a {\it T-flip-flop}, as shown by figure 9,
which provides both symbol and graph. For each pulse of the clock $CK$,
the preceding state and the input value $T$ determine a new state for the
memory, which can be known at the outside by means of the output $Q$.

\bigskip
\begin{center}
\begin{picture}(0,0)%
\psfig{file=ldiaf9.pstex}%
\end{picture}%
\setlength{\unitlength}{0.00087500in}%
\begingroup\makeatletter\ifx\SetFigFont\undefined
\def\x#1#2#3#4#5#6#7\relax{\def\x{#1#2#3#4#5#6}}%
\expandafter\x\fmtname xxxxxx\relax \def\y{splain}%
\ifx\x\y   
\gdef\SetFigFont#1#2#3{%
  \ifnum #1<17\tiny\else \ifnum #1<20\small\else
  \ifnum #1<24\normalsize\else \ifnum #1<29\large\else
  \ifnum #1<34\Large\else \ifnum #1<41\LARGE\else
     \huge\fi\fi\fi\fi\fi\fi
  \csname #3\endcsname}%
\else
\gdef\SetFigFont#1#2#3{\begingroup
  \count@#1\relax \ifnum 25<\count@\count@25\fi
  \def\x{\endgroup\@setsize\SetFigFont{#2pt}}%
  \expandafter\x
    \csname \romannumeral\the\count@ pt\expandafter\endcsname
    \csname @\romannumeral\the\count@ pt\endcsname
  \csname #3\endcsname}%
\fi
\fi\endgroup
\begin{picture}(3441,1710)(530,-1573)
\put(3556, 29){\makebox(0,0)[lb]{\smash{\SetFigFont{10}{12.0}{rm}$T=0$}}}
\put(2431,-1006){\makebox(0,0)[lb]{\smash{\SetFigFont{10}{12.0}{rm}0}}}
\put(3376,-466){\makebox(0,0)[lb]{\smash{\SetFigFont{10}{12.0}{rm}1}}}
\put(1981,-1546){\makebox(0,0)[lb]{\smash{\SetFigFont{10}{12.0}{rm}$T=0$}}}
\put(3286,-1096){\makebox(0,0)[lb]{\smash{\SetFigFont{10}{12.0}{rm}$T=1$}}}
\put(2161,-556){\makebox(0,0)[lb]{\smash{\SetFigFont{10}{12.0}{rm}$T=1$}}}
\put(677,-332){\makebox(0,0)[lb]{\smash{\SetFigFont{10}{12.0}{rm}$T$}}}
\put(1307,-332){\makebox(0,0)[lb]{\smash{\SetFigFont{10}{12.0}{rm}$Q$}}}
\put(1037,-1007){\makebox(0,0)[lb]{\smash{\SetFigFont{10}{12.0}{rm}$CK$}}}
\end{picture}

\bigskip
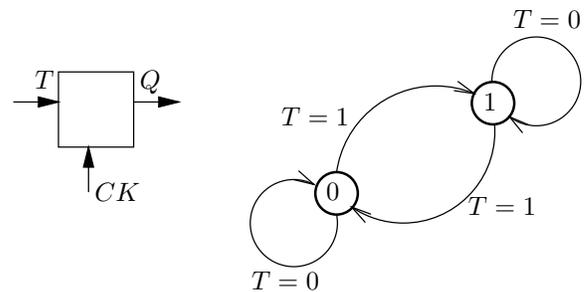
{\bf Figure 9.} {\it T-flip-flop}
\end{center}

\bigskip

If this flip-flop is built and tested separately, it can then be used as a
module (using two copies of it) for implementing the modulo four counter,
as shown by figure 10, where the state is binarily encoded on the
outputs of the two flip-flops $A$ and $B$ ($A$ being the least significant
digit).

The graph of this implement, represented on the right of figure 10,
is obtained by opening the connections at the $T$-input of the flip-flops.
It is the cartesian product of two graphs of $T$-flip-flops.
Each state transition is associated with a divergence and a convergence of
four arrows, which corresponds to the fact that the two flip-flops each possess
a $T$-input which can carry one information bit. The particular path followed
by
the modulo four counter in this graph reduces to the loop drawn as a bold line.
Arrows drawn as ordinary lines are never followed, once the connections between
the flip-flops have been settled. However, they should be counted when
evaluating the logical dissipation, since the flip-flops are tested
independently of their connections and the characteristic graph of each one
is not modified by their modular association.

One is faced here with two possible implement graphs for the same
function, defined by the graph of figure 8. The first implement (the
rotating wheel) has a graph identical to that of figure 8 and does not
dissipate. The second one (made of two $T$-flip-flops) has the graph of
figure 10. It is modular and dissipates.

\bigskip
\begin{center}
\begin{picture}(0,0)%
\psfig{file=ldiaf10.pstex}%
\end{picture}%
\setlength{\unitlength}{0.00087500in}%
\begingroup\makeatletter\ifx\SetFigFont\undefined
\def\x#1#2#3#4#5#6#7\relax{\def\x{#1#2#3#4#5#6}}%
\expandafter\x\fmtname xxxxxx\relax \def\y{splain}%
\ifx\x\y   
\gdef\SetFigFont#1#2#3{%
  \ifnum #1<17\tiny\else \ifnum #1<20\small\else
  \ifnum #1<24\normalsize\else \ifnum #1<29\large\else
  \ifnum #1<34\Large\else \ifnum #1<41\LARGE\else
     \huge\fi\fi\fi\fi\fi\fi
  \csname #3\endcsname}%
\else
\gdef\SetFigFont#1#2#3{\begingroup
  \count@#1\relax \ifnum 25<\count@\count@25\fi
  \def\x{\endgroup\@setsize\SetFigFont{#2pt}}%
  \expandafter\x
    \csname \romannumeral\the\count@ pt\expandafter\endcsname
    \csname @\romannumeral\the\count@ pt\endcsname
  \csname #3\endcsname}%
\fi
\fi\endgroup
\begin{picture}(3432,2316)(541,-2203)
\put(2791,-421){\makebox(0,0)[lb]{\smash{\SetFigFont{10}{12.0}{rm}B}}}
\put(3376,-61){\makebox(0,0)[lb]{\smash{\SetFigFont{10}{12.0}{rm}A}}}
\put(2566,-916){\makebox(0,0)[lb]{\smash{\SetFigFont{10}{12.0}{rm}00}}}
\put(3511,-916){\makebox(0,0)[lb]{\smash{\SetFigFont{10}{12.0}{rm}01}}}
\put(3511,-1816){\makebox(0,0)[lb]{\smash{\SetFigFont{10}{12.0}{rm}11}}}
\put(2566,-1816){\makebox(0,0)[lb]{\smash{\SetFigFont{10}{12.0}{rm}10}}}
\put(541,-1636){\makebox(0,0)[lb]{\smash{\SetFigFont{10}{12.0}{rm}1}}}
\put(991,-1636){\makebox(0,0)[lb]{\smash{\SetFigFont{10}{12.0}{rm}$T$}}}
\put(1261,-1636){\makebox(0,0)[lb]{\smash{\SetFigFont{10}{12.0}{rm}$Q$}}}
\put(1171,-2176){\makebox(0,0)[lb]{\smash{\SetFigFont{10}{12.0}{rm}$CK$}}}
\put(991,-646){\makebox(0,0)[lb]{\smash{\SetFigFont{10}{12.0}{rm}$T$}}}
\put(1261,-646){\makebox(0,0)[lb]{\smash{\SetFigFont{10}{12.0}{rm}$Q$}}}
\put(1846,-511){\makebox(0,0)[lb]{\smash{\SetFigFont{10}{12.0}{rm}B}}}
\put(1846,-1501){\makebox(0,0)[lb]{\smash{\SetFigFont{10}{12.0}{rm}A}}}
\put(766, 29){\makebox(0,0)[lb]{\smash{\SetFigFont{10}{12.0}{rm}Two}}}
\put(766,-196){\makebox(0,0)[lb]{\smash{\SetFigFont{10}{12.0}{rm}$T$-flip-flops
}}}
\end{picture}

\bigskip
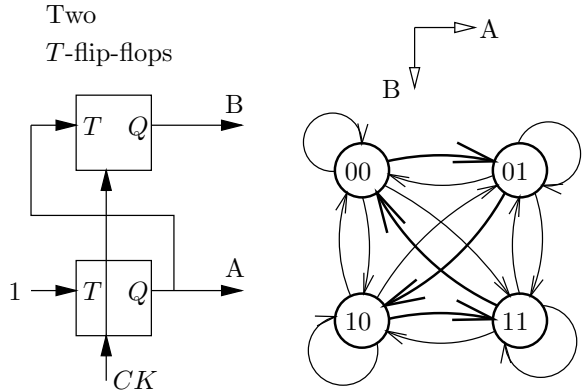
{\bf Figure 10.} {\it Two-bit counter}
\end{center}

\bigskip

\subsection{Modularity and extensivity property of
logical dissipation}

{\bf Definition}:
a physical implement of an automaton will be called {\bf modular},
if it consists of an association of subsets ({\it modules}) which are
physical implements of automata separately testable, associated through
connections which link the output of one module to inputs of other modules.

The association is not tested globally, as it is in general too complex.
However, each module is tested separately. It is the task of the designer
to prove that the modular association effectively realises the required
function, from the functions of the individual modules. This proof will not
be considered here. It could be obtained for instance through simulation
on a Turing machine.

In a modular association, the characteristic graph of each module is
independent of the other modules building the automaton: it is defined
by an independent test. The modular association does not modify
the characteristic graphs of the modules.

The graph of the modular implement is exhibited by the test. It is obtained
by opening the input connections of the modules, which means that the latter
are independently stimulated by the test machine. The graph of the implement
is thus the cartesian product of the graphs of the modules.
The global state is specified by the enumeration of the states of the modules.
The number of states of the graph of the implement is the product of the
numbers of states of the modules. The transition function of the implement
is the cartesian product of the transition functions of the modules.

Let us write the transition function for two modules $A$ and $B$
(generalisation to an arbitrary number of modules is immediate).
State, stimuli and transition functions will be denoted with the symbol
of each module in exponent. The transition function $G^{AB}$ of the
global implement is defined by:
$$Q^{AB}_{t+1} = (Q^{A}_{t+1}, Q^{B}_{t+1}) = G^{AB}(Q^{A}_t,
Q^{B}_t, S^{A}_t, S^{B}_t)$$
The fact that the transition function of each module is separately testable
(whatever the evolution of the other module) implies that the transition
function of the implement can be written:
$$G^{AB}(Q^{A}_t, Q^{B}_t, S^{A}_t, S^{B}_t) = (G^{A}(Q^{A}_t, S^{A}_t),
G^{B}(Q^{B}_t, S^{B}_t))$$
The number of arrows which leave state $(Q^A,Q^B)$ is equal to the product of
the numbers of arrows which leave states $Q^A$ and $Q^B$ in the graphs of the
separate modules. And for the arrows which reach a state alike.
Hence, the dissipation of the implement is the sum of the dissipations of
the modules. The logical dissipation thus appears as an {\bf extensive}
quantity.

Remark: unless being a simple juxtaposition of modules without communications,
a modular implement involves at least one module with an input which is
connected to the output of another module and which thus possesses at least
one divergence. As soon as this divergence is associated with a convergence,
the latter entails that the modular implement dissipates.

As an illustration, let us consider again the example of two $T$-flip-flops
implementing a modulo four counter. As shown by figure 10, the least
significant flip-flop
has its $T$-input connected to a constant.
A simplification would be to replace this flip-flop by a modulo two counter
whose symbol and graph are represented in figure 11.

\bigskip
\begin{center}
\begin{picture}(0,0)%
\psfig{file=ldiaf11.pstex}%
\end{picture}%
\setlength{\unitlength}{0.00087500in}%
\begingroup\makeatletter\ifx\SetFigFont\undefined
\def\x#1#2#3#4#5#6#7\relax{\def\x{#1#2#3#4#5#6}}%
\expandafter\x\fmtname xxxxxx\relax \def\y{splain}%
\ifx\x\y   
\gdef\SetFigFont#1#2#3{%
  \ifnum #1<17\tiny\else \ifnum #1<20\small\else
  \ifnum #1<24\normalsize\else \ifnum #1<29\large\else
  \ifnum #1<34\Large\else \ifnum #1<41\LARGE\else
     \huge\fi\fi\fi\fi\fi\fi
  \csname #3\endcsname}%
\else
\gdef\SetFigFont#1#2#3{\begingroup
  \count@#1\relax \ifnum 25<\count@\count@25\fi
  \def\x{\endgroup\@setsize\SetFigFont{#2pt}}%
  \expandafter\x
    \csname \romannumeral\the\count@ pt\expandafter\endcsname
    \csname @\romannumeral\the\count@ pt\endcsname
  \csname #3\endcsname}%
\fi
\fi\endgroup
\begin{picture}(2754,864)(620,-791)
\put(2251,-646){\makebox(0,0)[lb]{\smash{\SetFigFont{10}{12.0}{rm}0}}}
\put(3196,-106){\makebox(0,0)[lb]{\smash{\SetFigFont{10}{12.0}{rm}1}}}
\put(857,-737){\makebox(0,0)[lb]{\smash{\SetFigFont{10}{12.0}{rm}$CK$}}}
\put(1127,-62){\makebox(0,0)[lb]{\smash{\SetFigFont{10}{12.0}{rm}$Q$}}}
\end{picture}

\bigskip
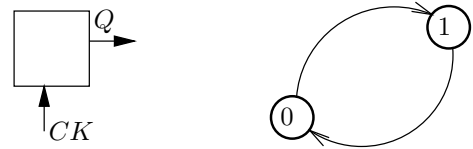
{\bf Figure 11.} {\it Modulo two counter}
\end{center}

\bigskip

The implement of the modulo four counter then becomes the one shown by
figure 12. In this modular association, the modulo two counter does not
dissipate, and the $T$-flip-flop is dissipative.
The association thus dissipates,
which results from the fact that one of the modules has an input,
which is necessary for transfering information from one module to the other,
if one is to realise a function less specialised than a simple juxtaposition
of non interacting modules.

\bigskip
\begin{center}
\begin{picture}(0,0)%
\psfig{file=ldiaf12.pstex}%
\end{picture}%
\setlength{\unitlength}{0.00087500in}%
\begingroup\makeatletter\ifx\SetFigFont\undefined
\def\x#1#2#3#4#5#6#7\relax{\def\x{#1#2#3#4#5#6}}%
\expandafter\x\fmtname xxxxxx\relax \def\y{splain}%
\ifx\x\y   
\gdef\SetFigFont#1#2#3{%
  \ifnum #1<17\tiny\else \ifnum #1<20\small\else
  \ifnum #1<24\normalsize\else \ifnum #1<29\large\else
  \ifnum #1<34\Large\else \ifnum #1<41\LARGE\else
     \huge\fi\fi\fi\fi\fi\fi
  \csname #3\endcsname}%
\else
\gdef\SetFigFont#1#2#3{\begingroup
  \count@#1\relax \ifnum 25<\count@\count@25\fi
  \def\x{\endgroup\@setsize\SetFigFont{#2pt}}%
  \expandafter\x
    \csname \romannumeral\the\count@ pt\expandafter\endcsname
    \csname @\romannumeral\the\count@ pt\endcsname
  \csname #3\endcsname}%
\fi
\fi\endgroup
\begin{picture}(3462,1992)(91,-1933)
\put(2341,-871){\makebox(0,0)[lb]{\smash{\SetFigFont{10}{12.0}{rm}00}}}
\put(3286,-871){\makebox(0,0)[lb]{\smash{\SetFigFont{10}{12.0}{rm}01}}}
\put(3286,-1771){\makebox(0,0)[lb]{\smash{\SetFigFont{10}{12.0}{rm}11}}}
\put(2341,-1771){\makebox(0,0)[lb]{\smash{\SetFigFont{10}{12.0}{rm}10}}}
\put(2566,-466){\makebox(0,0)[lb]{\smash{\SetFigFont{10}{12.0}{rm}B}}}
\put(3151,-106){\makebox(0,0)[lb]{\smash{\SetFigFont{10}{12.0}{rm}A}}}
\put( 91,-1546){\makebox(0,0)[lb]{\smash{\SetFigFont{10}{12.0}{rm}counter}}}
\put( 91,-1321){\makebox(0,0)[lb]{\smash{\SetFigFont{10}{12.0}{rm}Modulo two}}}
\put(226,-61){\makebox(0,0)[lb]{\smash{\SetFigFont{10}{12.0}{rm}$T$-flip-flop}}}
\put(1171,-1906){\makebox(0,0)[lb]{\smash{\SetFigFont{10}{12.0}{rm}$CK$}}}
\put(1261,-1366){\makebox(0,0)[lb]{\smash{\SetFigFont{10}{12.0}{rm}$Q$}}}
\put(1846,-1231){\makebox(0,0)[lb]{\smash{\SetFigFont{10}{12.0}{rm}A}}}
\put(1846,-241){\makebox(0,0)[lb]{\smash{\SetFigFont{10}{12.0}{rm}B}}}
\put(991,-376){\makebox(0,0)[lb]{\smash{\SetFigFont{10}{12.0}{rm}$T$}}}
\put(1261,-376){\makebox(0,0)[lb]{\smash{\SetFigFont{10}{12.0}{rm}$Q$}}}
\end{picture}

\bigskip
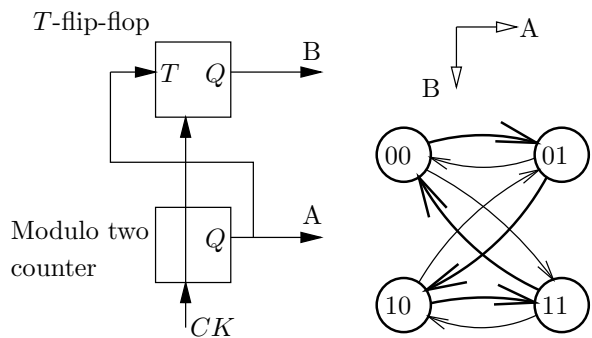
{\bf Figure 12.} {\it One of the modules does not dissipate}
\end{center}

\bigskip

\subsection{Cost of a modular implement}

With the definition previously given for the cost, it is easily seen that the
cost of a modular implement is smaller than the one of a non-modular
implement (and in an exponential way). Indeed, the complexity of
the test for a modular implement is the sum of the complexities of the tests
of the modules and does not grow any more like the number of states of the
global implement.

Modularity allows to decrease specialisation in the structure of the
implement graph, but this entails an increase in the number of arrows and
of convergences, and hence dissipation.

\section{Dissipation of computation}

In order to discuss dissipation of {\it computation} in a general way,
one can consider Turing machines \cite{TUR 36}.
Only {\it deterministic} Turing machines will be considered here, and their
physical implementation by means of automata whose graph is stable in time
and testable, which allows to define a date of test (see section III).

To determine whether implements of Turing machines must dissipate,
one may ask if there can exist a physical implement which is,
from a logical point of view, globally equivalent to a Turing machine
but which does not dissipate.

Turing's logical description contains parts which communicate:
The {\it tape} and the {\it head} (we shall denote by "head" all which
is not a tape in the machine). This corresponds to several finite automata:

- one automaton in the head of the machine,

- a juxtaposition of an infinite number of memory cells in the tape(s).
Each cell of a tape is a finite automaton, whose number of states is at least
equal to the number of symbols in the alphabet.

Two finite automata (the head and one memory cell) communicate during
read/write operations of the head on the tape, and change their states
possibly together.

If the implement of the machine strictly follows this description,
that is if it is an assembly of modules which reproduce the functions of
these parts and which can be tested separately, one is in the situation of
a modular implementation, and there will be two causes for dissipation:

- the automaton of a memory cell is finite, although the number of write
operations must not be bounded. Hence, writing a symbol periodically
meets a convergence which entails a loss of previously memorized information.

- the head automaton can involve convergences in its graph, which is at
least the case for universal Turing machines, according to the following
lemma.

{\bf Lemma}: the graph of the head automaton of a universal Turing machine
contains
at least one convergence.

{\bf Demonstration}: a universal Turing machine can simulate any Turing
machine,
in particular one which is engaged in a computation which does not halt.
As its head automaton is finite, it contains at least one loop.
In order not to involve a convergence, this loop should contain all the states
of the automaton. Its evolution would then be periodic.
But there exist computations which are infinite and non periodic.
This non-periodicity for a finite automaton implies the existence of
at least one convergence in its graph.

Is modularity necessary?

If one wants to preserve the {\it programmability} property,
that is if one wants
the machine to be testable independently of the program (without testing all
possible programs), then one has to implement the head and the tape
as two independent modules. One can conclude that the implement
of a universal Turing machine is necessarily modular, and that it dissipates
proportionally  to the number of computation steps.

\subsection{Finite implements of specialised Turing machines}

In practice, implements of Turing machines are finite, that is their tape
is of finite length. Of course, this finiteness sets a bound on the
complexity of allowed computations, but all computations with a complexity
which is smaller than a previously fixed bound can be performed,
as it is usually
done with computers which are also finite machines.

Within this framework, programmability still leads to modularity and
dissipation. But if one considers machines which are specialised for a unique
computation, one can then ask whether a particular computation which halts
can be implemented with a non-dissipative automaton (possibly with a
non-modular implement).

If one considers a particular program (a particular INPUT tape) leading
to a computation which halts, then the assembly head + tape is globally
equivalent to a linear non-dissipative automaton as the one of figure 13.
Indeed, as the tape lies {\it inside}, it is a finite inputless automaton
which never goes twice through the same state (otherwise it would
indefinitely come back to this state and computation would not halt).
It is thus clear that any halting computation can be associated with
a non-dissipative graph.

\bigskip
\begin{center}
\begin{picture}(0,0)%
\psfig{file=ldiaf13.pstex}%
\end{picture}%
\setlength{\unitlength}{0.00087500in}%
\begingroup\makeatletter\ifx\SetFigFont\undefined
\def\x#1#2#3#4#5#6#7\relax{\def\x{#1#2#3#4#5#6}}%
\expandafter\x\fmtname xxxxxx\relax \def\y{splain}%
\ifx\x\y   
\gdef\SetFigFont#1#2#3{%
  \ifnum #1<17\tiny\else \ifnum #1<20\small\else
  \ifnum #1<24\normalsize\else \ifnum #1<29\large\else
  \ifnum #1<34\Large\else \ifnum #1<41\LARGE\else
     \huge\fi\fi\fi\fi\fi\fi
  \csname #3\endcsname}%
\else
\gdef\SetFigFont#1#2#3{\begingroup
  \count@#1\relax \ifnum 25<\count@\count@25\fi
  \def\x{\endgroup\@setsize\SetFigFont{#2pt}}%
  \expandafter\x
    \csname \romannumeral\the\count@ pt\expandafter\endcsname
    \csname @\romannumeral\the\count@ pt\endcsname
  \csname #3\endcsname}%
\fi
\fi\endgroup
\begin{picture}(3202,1146)(541,-1303)
\put(541,-241){\makebox(0,0)[lb]{\smash{\SetFigFont{10}{12.0}{rm}Start}}}
\put(1216,-1276){\makebox(0,0)[lb]{\smash{\SetFigFont{10}{12.0}{rm}Stop}}}
\end{picture}

\bigskip
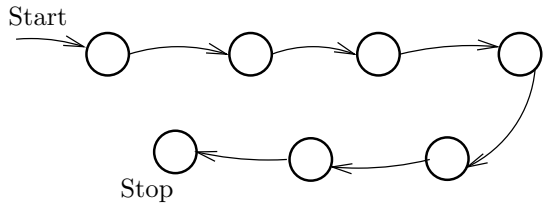
{\bf Figure 13.} {\it Linear automaton}
\end{center}

\bigskip

But the necessity for modularity is then entailed by a size
argument on the number of states: a Turing machine including a tape
able to memorize 100 bits contains $2^{100}$ states (to be multiplied
by the number of states in the head), which already exceeds the time of tests
one can envisage. Whilst it is a very small size for a memory.

Thus, even for a finite machine, even for a machine which only
performs a single computation (hence specialised and inputless),
modularity is necessary if the parts are not to exceed a reasonable
size for the number of states.

\subsection{Bennett's reversible machine}

Bennett has proposed a machine whose dissipation is not linked to computation
time and which can be built as follows: with any Turing machine $\cal S$
(that is with any head automaton of a Turing machine), one can associate
a reversible Turing machine $\cal R$ which makes
any working tape evolve in the same way as
$\cal S$, but which further memorizes the computation history
and then uses the latter to {\it undo} the computation, the whole
being deterministic and reversible \cite{BEN 73}.
Other examples of simple reversible Turing machines have been proposed since
then (see for instance \cite{MOR 89}).
The question is that of logical reversibility for the global computation
(Bennett writes "the whole state machine"). Bennett's proof exhibits for the
machine $\cal R$ a head automaton which can be made explicit before computation
(which only depends on the head automaton of the machine $\cal S$ and not on
the INPUT tape) and which is able to engage in any computation (on any
tape provided after the date of test of the head).

But Bennett's conclusions on the logical reversibility of the global machine
(head + tapes) do not explicit a reversible graph for this machine,
which would be independent of the INPUT tape and would allow an implementation
which would be {\it testable before computation}. Indeed, only the graph
of the head automaton is made explicit before computation, but only
the {\it global machine} (that is the global computation) is reversible.
The graph of the global machine is specialised and depends on the
INPUT tape. It can be made explicit, but this is achieved by the computation
itself (if it halts). Hence, it is only known at the end of the computation
and cannot be the graph for an implement which is testable before the beginning
of computation.

The logical reversibility of the global computation must not make one
believe in the possibility to implement a non-dissipative machine by
reproducing Bennett's logical structure, as the division head/tape
would necessarily imply modularity and hence a dissipation proportional
to computation time.

The graphs of the modules taken separately include convergences although
that of the global machine (global computation) includes none. This
is understood by recalling that useful information is permuted within
head and tapes and is never lost for the global machine, although it is
sometimes lost for some of the modules.

The global machine is a specialised machine which depends on the INPUT tape.
When computation has halted,
its graph has become explicit. It is a linear graph,
like the one of figure 13, whose number of states is equal to the number of
steps of the machine $\cal R$, that is $4n + 4r + 5$ where $n$ is the
number of steps of the machine $\cal S$, that is the computation complexity
in time, and $r$ is the number of symbols of the result.

Thus, Bennetts' construction does not allow to implement a universal
non-dissipative machine, but leads to construct as many reversible machines
as distinct computations. The physical implementation of such machines
remains, for testability reasons, very strongly limited in complexity.

\section{Discussions}

\subsection{Logical separation}

The definition of the physical implement of an automaton through its test
implicitly contains the idea that a {\it logical separation} can be made
between the inside and the outside of the automaton, and that exchanges
of information between inside and outside (in both directions) must follow
the channels defined by inputs and outputs. This means that one should exclude
the possibility of correlations between the physical internal state of
the automaton and its environment, which would have a logical meaning
but would not be linked to exchanges through inputs/outputs. Physical
correlations can nonetheless exist, as long as they are not involved
in the logical function.

Separation between inside and outside is not necessarily the isolation
of a connected part of space, which would be limited by the "walls"
of the black box, but amounts to functionally isolate a physical system,
part of whose coordinates define the state of the automaton, the latter
knowing its environment only by means of the information present on its inputs.
By definition, such an automaton has no knowledge of the origin of the
information  present on its inputs, neither of the use of its ouputs.
Furthermore, its {\it logical dynamics} are independent of the nature of its
outside connections. This separation is not a consequence of physical laws,
but is on the contrary a necessary logical assumption in order to
envisage the faithfulness and testability of implements.

The part played by logic in physics cannot be reduced to the distinction
of some physical systems for realising logical functions. Automata are also
necessary elements for studying physical systems. Indeed, the description
of physical systems requires to perform particular interactions with them,
in order to obtain information on their state, called {\it measurements}.
By definition, the information provided by the measurement must allow
to be memorized and treated by logical systems. Measurements involve
automata, which realise the necessary interface between physical quantum
observables and the observed values which undergo a classical
logical treatment \cite{WIT 63}. This characteristic role played by the
automaton is particularly illustrated by Maxwell's demons.

Logical separation and modularity are fundamental properties of physical
implements of automata, imposed by testability. They have important
consequences for the physical constitution of automata. They imply that
in a modular implement, the internal connections of the automaton which are
external connections of modules (relating the output of one module
to the input of another module of the same implement) only carry
information which can be tested classically. That is, a modular implement
can only reproduce {\it classical} sequential functions. In other words,
the non-classical realisations which one can envisage (for instance
exploiting quantum properties \cite{DEU 85}) are necessarily non-modular.
Their complexity is then limited by testability constraints.

\subsection{Description of automata by means of recursive functions}

We have considered that an automaton is defined by a finite list of states
and transitions and that this list is explicitly given. This restriction
is made necessary for physically constructing and testing the implement.

But from a strictly formal point of view, one can wonder whether one can
envisage to describe  the graph by means of a recursive function.
It might then be difficult to say whether the graph is finite by
simply inspecting its description formula. If the graph is the result
of a computation, undecidability of halting of the computation will
translate into undecidability of the structure and finiteness of the graph.
If the graph computation halts, this computation has for consequence to
make the graph explicit and thus to allow to construct and test the
implement.

A more delicate situation would arise if the implement is built during
computation with the help of another automaton (a robot) which puts components
together following the instructions of a program. This has been excluded by
our definition which imposes testing before using. But such a possibility
could allow another form of existence for Bennett's reversible machine.
Indeed, although it is not explicit before computation, its construction
algorithm is known (from the INPUT tape and the head specified by Bennett)
and it is constructed by the computation. In that case, the analysis
of dissipation cannot be limited to counting divergences and converges in
the final graph, but must also take into account the graph of the robot
assembling the components. Preceding discussions can then be applied to
this whole set.

\section{Conclusion}

We have given, for the implement of a finite automaton,
a definition of the logical dissipation
which is a function of its graph.

We have defined the modularity of automata implementation and shown that
dissipation is linked to that modularity. A computing machine dissipates
proportionally to computation time if the machine is programmable
or of a reasonable size, since it is then necessarily modular.
Otherwise, the specialisation of a non-dissipative machine or the complexity of
its test make it incompatible with the universality underlying the notion
of computation.

Modularity diminishes specialisation. It allows to reach more complex
and more varied computations with a more simple test. As a counterpart,
it entails dissipation. Turing machines can be both universal and simple.
On another hand, Bennett's reversible machine is specialised or complex to
test.
Dissipation appears, through modularity, as the property which allows simple
machines to perform varied computations of an arbitrary complexity.

We have not discussed the physical mechanisms which can lead to dissipation,
neither, {\it a fortiori} tried to specify the value of the conversion
coefficient between logical and physical dissipation. Such a discussion
should be parallelled with a discussion of physical limitations in
the registration of data, in their transmission and their conservation
(limits related to measurement sensitivity, causality and stability).
We have seen that logical dissipation is a necessary condition for
implementing some automata. Conversely, the role played by automata in
measurement, hence in the test, shows the close link between physical
dissipation and the {\it logical separation} necessary to the logical
functioning of physical implements of automata.

\section{Acknowledgements}

We thank G. Cousineau, W. Mercouroff and J. Stern, and T.S.I. referees for
helping us to bring this text to its final form.

\end{document}